\documentclass[aps,prl,twocolumn,showpacs,amsmath,amssymb,superscriptaddress]{revtex4-2}
\usepackage{braket}%Dirac Notation in QM
\usepackage{cases}
\usepackage{color}
\usepackage{graphicx,subfigure}% Include figure files
\usepackage{float}
\usepackage{amssymb}
\usepackage{amsthm}

\usepackage{bm}% bold math
\usepackage{hyperref}
\hypersetup{
    colorlinks=true, 
    linkcolor=cyan,
    citecolor=magenta, 
    filecolor=magenta, 
    urlcolor=cyan,
    runcolor=cyan
}
\usepackage{array}
\usepackage[capitalise]{cleveref}

\usepackage{commath}
\usepackage{color}
\usepackage{amsmath}
\usepackage{adjustbox}
\usepackage[normalem]{ulem}
\def\be{\begin{equation}}
\def\ee{\end{equation}}

\newcommand{\bes} {\begin{subequations}}
\newcommand{\ees} {\end{subequations}}
\newcommand{\beq}{\begin{equation}}
\newcommand{\eeq}{\end{equation}}

\def\>{\rangle}
\def\<{\langle}
\def\Tr{\mathrm{Tr}}

\newcommand{\ketb}[2]{|{#1}\>\!\<#2|}

\begin{document}

\title{Qudit Dynamical Decoupling on a Superconducting Quantum Processor}

\author{Vinay Tripathi}
\thanks{These two authors contributed equally to this work.}
\affiliation{Department of Physics \& Astronomy, University of Southern California,
Los Angeles, CA 90089, USA}
\affiliation{Center for Quantum Information Science \& Technology, University of
Southern California, Los Angeles, CA 90089, USA}
\author{Noah Goss}
\thanks{These two authors contributed equally to this work.}
\affiliation{Quantum Nanoelectronics Laboratory, Department of Physics, University of California at Berkeley, Berkeley, CA 94720, USA}
    \affiliation{Applied Math and Computational Research Division, Lawrence Berkeley National Lab, Berkeley, CA 94720, USA}
\author{Arian Vezvaee}
\affiliation{Center for Quantum Information Science \& Technology, University of
Southern California, Los Angeles, CA 90089, USA}
\affiliation{Department of Electrical \& Computer  Engineering, University of Southern California, Los Angeles, California 90089, USA}
\author{\\Long B. Nguyen}
\affiliation{Quantum Nanoelectronics Laboratory, Department of Physics, University of California at Berkeley, Berkeley, CA 94720, USA}
    \affiliation{Applied Math and Computational Research Division, Lawrence Berkeley National Lab, Berkeley, CA 94720, USA}
\author{Irfan Siddiqi}
\affiliation{Quantum Nanoelectronics Laboratory, Department of Physics, University of California at Berkeley, Berkeley, CA 94720, USA}
    \affiliation{Applied Math and Computational Research Division, Lawrence Berkeley National Lab, Berkeley, CA 94720, USA}
\author{Daniel A. Lidar}
\affiliation{Department of Physics \& Astronomy, University of Southern California,
Los Angeles, CA 90089, USA}
\affiliation{Center for Quantum Information Science \& Technology, University of
Southern California, Los Angeles, CA 90089, USA}
\affiliation{Department of Electrical \& Computer  Engineering, University of Southern California, Los Angeles, California 90089, USA}
\affiliation{Department of Chemistry, University of Southern California, Los Angeles, California 90089, USA}

%\date{\today}

\begin{abstract}

Multi-level qudit systems are increasingly being explored as alternatives to traditional qubit systems due to their denser information storage and processing potential. However, qudits are more susceptible to decoherence than qubits due to increased loss channels, noise sensitivity, and crosstalk.  To address these challenges, we develop protocols for dynamical decoupling (DD) of qudit systems based on the Heisenberg-Weyl group. We implement and experimentally verify these DD protocols on a superconducting transmon processor that supports qudit operation based on qutrits $(d=3)$ and ququarts $(d=4)$. Specifically, we demonstrate single-qudit DD sequences to decouple qutrits and ququarts from system-bath-induced decoherence. We also introduce two-qudit DD sequences designed to suppress the detrimental cross-Kerr couplings between coupled qudits. This allows us to demonstrate a significant improvement in the fidelity of time-evolved qutrit Bell states. Our results highlight the utility of leveraging DD to enable scalable qudit-based quantum computing.
\end{abstract}

\maketitle

Multilevel quantum systems, also known as qudits~\cite{Gottesman:99}, offer potentially superior computational capabilities and denser information encoding relative to traditional qubit-based schemes~\cite{Muthukrishnan:2000aa, bartlett2002quantum, Grace2006aa, kues2017chip, erhard2018experimental, PhysRevLett.123.070505, davis2019photon, Low2020PRR, Blok2021, ringbauer2022universal, Goss2022, Srivastav2022, chi2022programmable, Liu2023PRX, hrmo2023native, Fischer2023PRXQ, Raissi2024PRXQ, nguyen2023empowering, roy2024synthetic, vezvaee2024quantum}. In addition, qudits enable resource-efficient fault-tolerant quantum computation~\cite{Campbell2012PRX, Campbell2014PRL,Majumdar2018} and the exploration of complex novel quantum applications~\cite{Fedorov2011Nature, Bookatz:2014aa, Gokhale2019, nguyen2022programmable, kiktenko2023realization} with reduced resource requirements. However, in superconducting devices, qudits are more susceptible to low-frequency noise and correlated errors, which pose significant challenges~\cite{goss2023extending}. Addressing these requires the development of scalable strategies for the mitigation and suppression of decoherence~\cite{Lidar-Brun:book}. Dynamical decoupling (DD)~\cite{Viola1998PRA, Zanardi:1999fk, Viola1999PRL,Duan:98e,Vitali1999,Suter:2016aa} is a powerful technique designed to enhance the fidelity of quantum states by employing carefully timed control pulses. It has been used to effectively decouple superconducting qubits from environmental noise~\cite{Pokharel2018, Ezzell2023, Baumer2024} and unwanted crosstalk~\cite{Tripathi2022,Zhou2023, Baumer2023, seif2024suppressing}. DD has been studied across a broad spectrum of qubit-based systems, but its experimental application to qudits has been limited primarily to trapped ions and nitrogen-vacancy ensembles~\cite{Yuan2022, zhou2023robust,Napolitano2021PRR,VitanovPRA2015}, and very recently to enhance the fidelity of a qutrit-assisted three-qubit Toffoli gate on an IBM transmon device~\cite{iiyama2024qudit}.
\begin{figure}[h]
\centering
\includegraphics[width=0.8\linewidth]{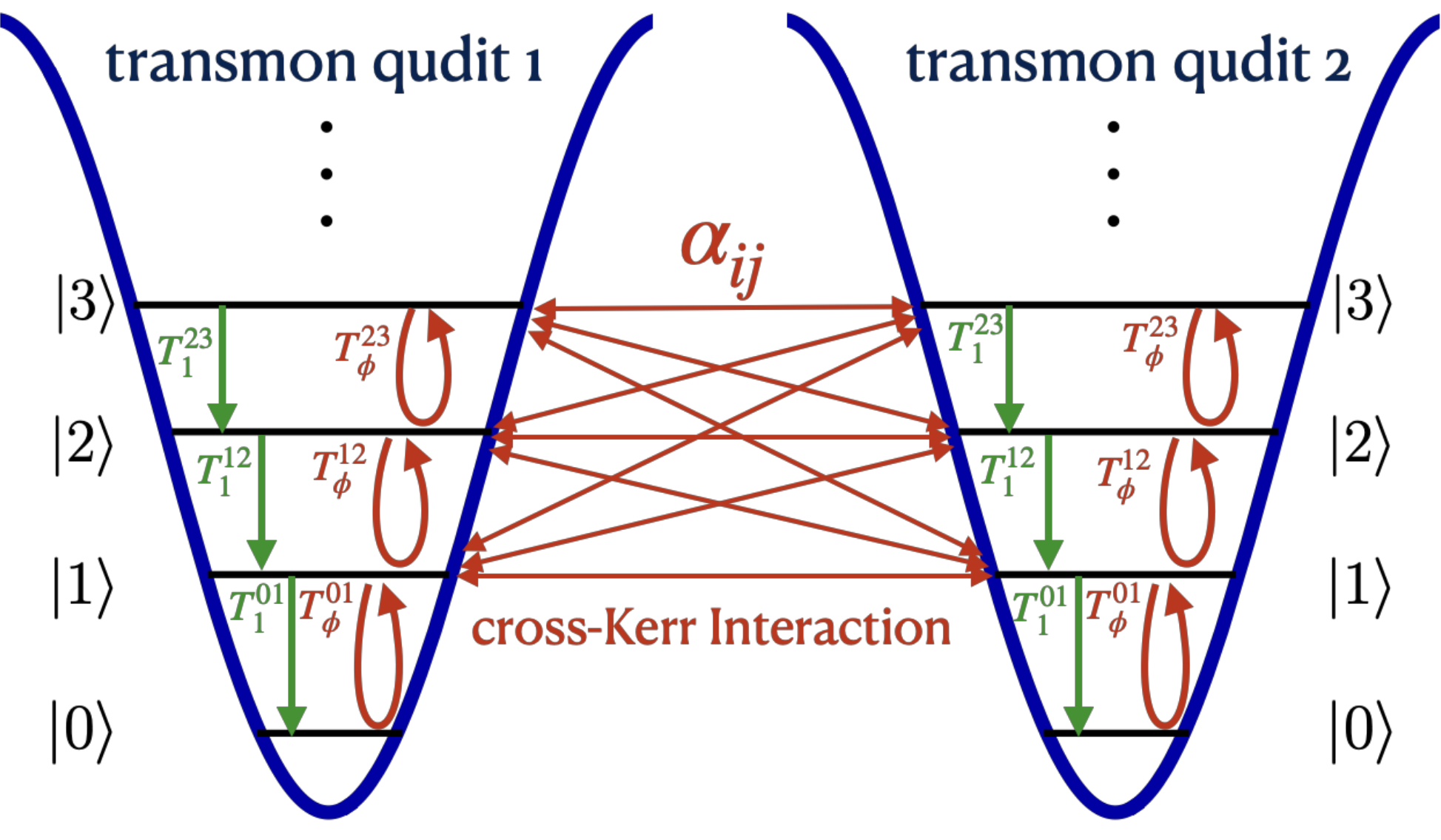}
\caption{%Top: 
Schematic illustration of two transmon qudits with quantized energy levels affected by relaxation and dephasing errors, along with the qudit-qudit cross-Kerr couplings $\alpha_{ij}$.}
%Bottom: Schematic of a concatenated DD sequence applied to the two qudits, which suppresses all the cross-Kerr interactions as well as dephasing channels.}
\label{fig:pulse}
\end{figure}

\begin{figure*}[t]
\centering
\includegraphics[width=.45\linewidth]{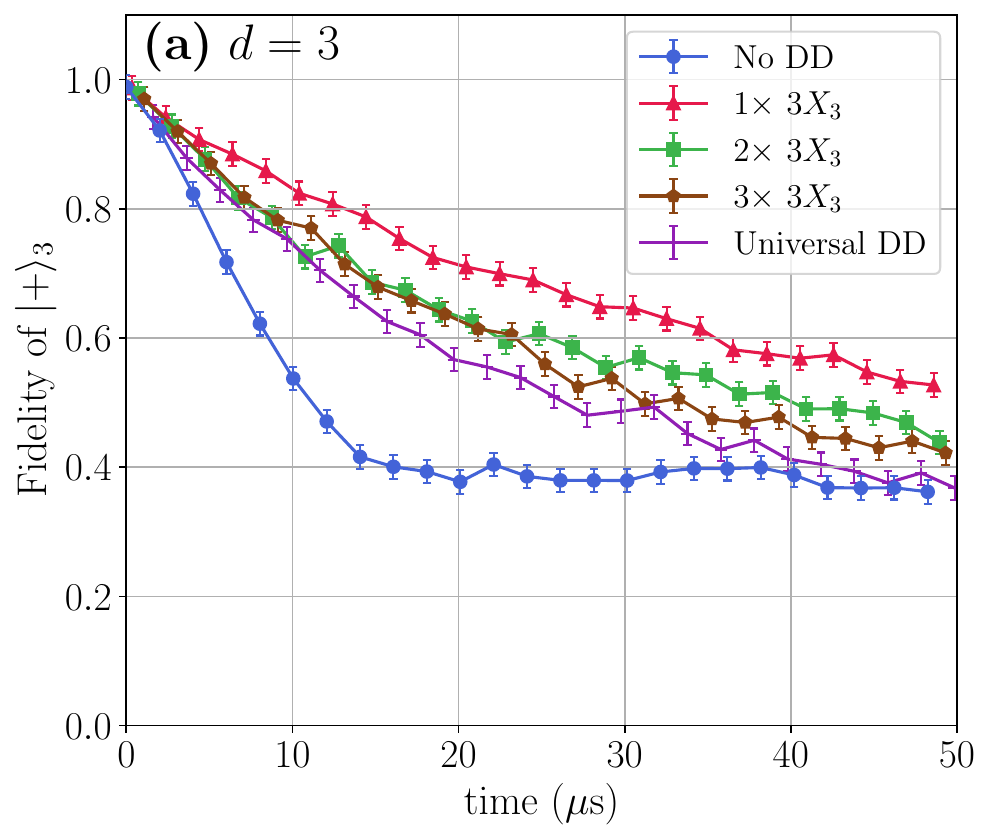}
\includegraphics[width=.45\linewidth]{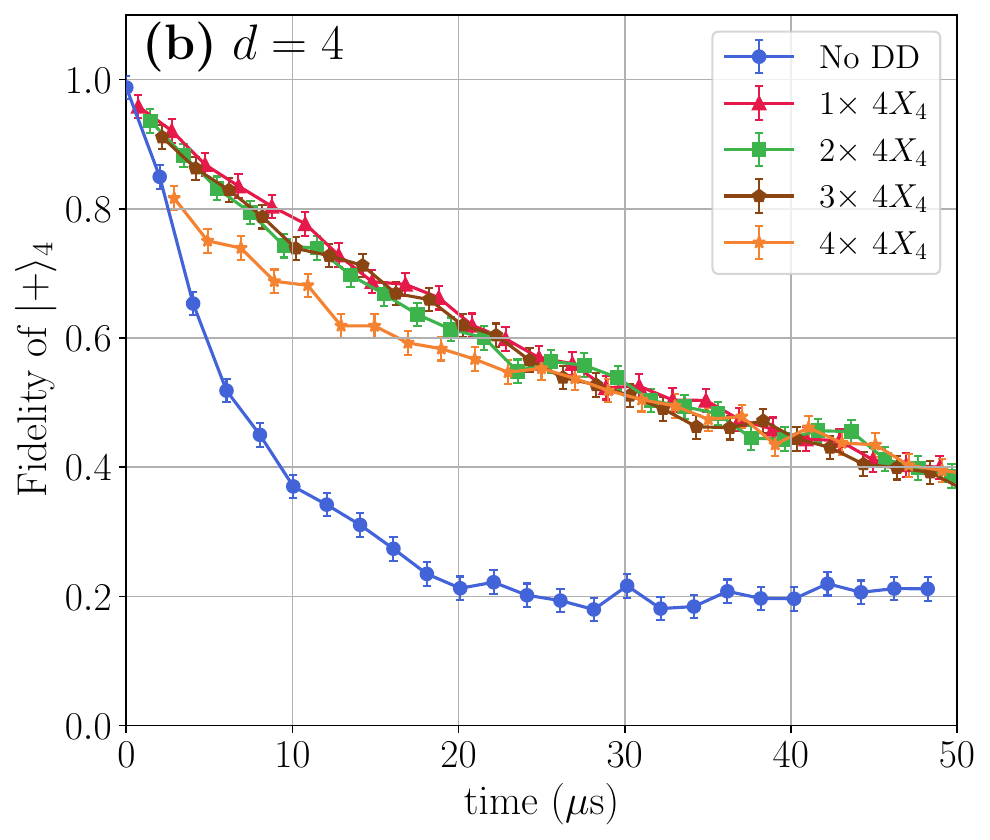}
\caption{Experimental results showing the fidelity of the qudit uniform superposition state $|+\rangle_d$ as a function of total time under free evolution (No DD) and $dX_D$, for (a) qutrits and (b) ququarts. The minimum pulse interval is $\tau_{\min} = 120$ ns. 
For qutrits, we implement $1$, $2$, and $3$ repetitions of the $3X_3$ sequence, with corresponding pulse intervals $3\tau$, $3\tau/2$, and $\tau$, respectively. The total evolution time is always $9\tau$. Universal DD is a sequence of $9$ pulses formed by cycling over the entire HWG, applied once with a pulse interval of $\tau$ and $\tau_{\min} = 180$ ns. 
For ququarts, we similarly implement $1$, $2$, $3$, and $4$ repetitions of the $4X_4$ sequence, with corresponding pulse intervals of $4\tau$, $2\tau$, $\frac{4}{3}\tau$, and $\tau$ respectively, where the total time is always $16\tau$.}
\label{fig:single_qutrit}
\end{figure*}

In this work, we present a general DD framework tailored for \textit{qudits} and experimentally demonstrate its effectiveness using coupled superconducting transmon circuits \cite{Koch2007} operated as qutrits $(d=3)$ and ququarts $(d=4)$.  Our framework employs the Heisenberg-Weyl (HW) group, which has found many uses in the study of $d$-dimensional quantum systems~\cite{Vourdas_2004, Wang2020Review}. We present DD sequences for universal noise suppression and also introduce a single-axis DD sequence designed to suppress the prevalent $1/f$ dephasing noise that plagues superconducting qudits. We then introduce a multi-qudit DD sequence designed to suppress unwanted cross-Kerr interactions between coupled qudits (see \cref{fig:pulse}), which stand in the way of scaling superconducting qudit systems~\cite{Blok2021, goss2023extending}. 
Using our DD sequences, we additionally report a significant enhancement in preserving the fidelity of a qutrit Bell state over time. This work serves as a proof-of-concept demonstration of the efficacy and scalability of active refocusing techniques in qudit systems and provides a stepping stone toward operating large-scale high-dimensional architectures.

\textit{Qudit dynamical decoupling theory}.---
Building on the general symmetrization ideas of Refs.~\cite{Zanardi:1999fk,Viola1999PRL}, the theory of qudit DD was developed in Refs.~\cite{Wocjan2006, rotteler2006equivalence, Rotteler_Wocjan_2013, Bookatz_2016}. We briefly introduce essential terminology and present a detailed review in the Supplementary Materials (SM), where we also generalize the theory. 

The decoupling group $\mathcal{G}_d$ is a set of unitary transformations (pulses) $g_j$ acting purely on the system: $\mathcal{G}_d=\left\{g_0, \cdots, g_K\right\}$, where $g_0$ is the $d$-dimensional identity operator $I$.  Under the instantaneous and ideal pulse assumptions, cycling over all elements of the group yields the following DD pulse sequence~\cite{Zanardi:1999fk,Viola1999PRL}: 

\begin{equation} 
U(T)=
\prod_{j=0}^K g_j^{\dagger} f_\tau g_j .
\label{eq:cycle}
\end{equation}
Here, $\tau$ is pulse interval, $T=|\mathcal{G}_d|\tau = (K+1)\tau$ is the total time taken by the sequence, and $f_\tau = e^{-i\tau H}$ is the free-evolution unitary, where $H$ is the total Hamiltonian of the system and the bath. A universal DD sequence for a qubit ($d=2$) is obtained by choosing the decoupling group as the Pauli group $\mathcal{G}_2=\{I,X,Y,Z\}$, for which $U(T)$ simplifies into the well-known XY4 sequence $U(4\tau) = Y f_\tau X f_\tau Y f_\tau X f_\tau$~\cite{Maudsley:1986ty}.

For $d>2$, we instead use the decoupling group to be the Heisenberg-Weyl group (HWG) of order $d^2$, which generalizes the Pauli group. The HWG is generated by the following shift and phase operators:
\begin{equation} 
    X_d\equiv \sum_{k=0}^{d-1} \ketb{(k+1)\bmod d}{k}, \, 
    \quad Z_d\equiv \sum_{k=0}^{d-1} \gamma_d^k \ketb{k}{k},
\label{eq:fund-ops}
\end{equation}
where $\gamma_d=e^{2\pi i/d}$ is the $d$th root of unity. The remaining HWG elements are given by $\Lambda_{\alpha\beta}=(-\sqrt{\gamma_d})^{\alpha\beta}X_d^\alpha Z_d^\beta$ where $\alpha,\beta\in \mathbb{Z}_d = \{0,1,2,...,d-1\}$. 

The dominant decoherence mechanism in transmon qutrits and ququarts is dephasing due to $1/f$ noise \cite{Tripathi2023}, which has been connected to charge fluctuations and higher level charge sensitivity \cite{Koch2007, PhysRevB.104.224509}. 
Thus, for single qudits, we focus on single-axis DD sequences consisting only of the shift operator and its powers, i.e., the decoupling group formed by the HW subgroup $\{X_d^k\}_{k=0}^{d-1}$. Note that $(X_d^k)^\dagger = X_d^{d-k}$. Thus, cycling over these operators, we obtain $U(T) = (X_d^{1} f_\tau X_d^{d-1})...(X_d^{d-2} f_\tau X_d^2)(X_d^{d-1} f_\tau X_d)(I f_\tau I)$. Simplifying, this becomes the sequence $dX_d \equiv X_d f_\tau X_d f_\tau X_d ... X_d f_\tau$.

\textit{Single qudit $dX_d$ experiment}.---
We conduct all our experiments on a superconducting transmon qudit processor with $d=3$ and $4$; other parameters are detailed in the SM. Since DD sequences are particularly effective against low-frequency noise~\cite{ShiokawaLidar:02}, and superconducting circuits are especially susceptible to such noise when higher excited states are targeted~\cite{Blok2021}, we focus primarily on the $dX_d$ sequence family. The underlying cycle operator $X_d$ is compiled using $2(d-1)$ native $\sqrt{\sigma^x_s}$ subspace rotations where $s\in \{(0,1),(1,2),..,(d-1,d)\}$, and $\sigma^x_{(i,j)} = \ketb{i}{j}+\ketb{j}{i}$ is the Pauli-$x$ operator between levels $i$ and $j$. 
\cref{fig:single_qutrit} presents our experimental single-qudit $dX_d$ results. Free evolution (no DD) corresponds to the preparation of a uniform qudit superposition state $|+\rangle_d \equiv (|0\rangle + \dots + |d-1\rangle)/\sqrt{d}$, waiting for a specified delay time, unpreparing the state, and finally measuring the qudit. Assuming ideal preparation, unpreparation, and measurement, the fidelity of the superposition state $|+\rangle_d$ is the probability of finding the qudit back in the $|0\rangle$ state. We then repeat the experiment with the $dX_d$ sequence applied during the delay time and study its impact on the state fidelity.  

\cref{fig:single_qutrit}(a) presents the results for the qutrit experiments. Crucially, all the DD curves exhibit an improvement over the free evolution (No DD) experiment, confirming the effectiveness of our DD sequences in suppressing decoherence. In more detail, for each total time $T=9\tau$ we conducted a free evolution experiment and four DD experiments: $1$, $2$, or $3$ repetitions of $3X_3$, and universal qutrit DD (the full order-$9$ HWG), with respective pulse intervals of $\tau_1 = 3\tau$, $\tau_2 = \frac{3}{2}\tau$, $\tau_3 = \tau$, and $\tau_{\text{univ.}} = \tau$. DD theory predicts that for instantaneous, ideal pulses, state preservation fidelity increases monotonically as the pulse interval decreases for a fixed total evolution time~\cite{Khodjasteh:2007zr,UL:10}. Moreover, universal DD is expected to outperform single-axis DD. Our results exhibit the opposite of both expectations: the single repetition $1\times 3X_3$ experiment, with the longest pulse interval $\tau_1=3\tau$, yields the highest fidelity, while the universal sequence, with the shortest interval $\tau$ (as for $3\times 3X_3$) yields the lowest DD fidelity. The reason for these results is likely to be the presence of coherent pulse errors, which accumulate more detrimentally the longer the pulse sequence, and whose effect overwhelms the benefit of shorter pulse intervals~\cite{Souza:2011aa,Peng:2011ly}. While $X_3$ gates can be decomposed in terms of four native $\sqrt{\sigma^x_s}$ gates, the remaining HW pulses require six native $\sqrt{\sigma^x_s}$ gates, so that $\tau_{\min} = 180$ ns for universal DD compared to $\tau_{\min} = 120$ ns for the $3X_3$ sequences. This additional opportunity for the accumulation of coherent errors explains why the universal sequence underperforms the $3\times 3X_3$ sequence. The superior performance of the $3X_3$ sequences also confirms that the dominant source of noise is dephasing. 

Similar improvements are observed with DD for ququarts, as shown in \cref{fig:single_qutrit}(b). Since ququarts are more susceptible to charge noise due to the involvement of the third excited state~\cite{Koch2007}, the free evolution fidelity is significantly lower than in the qutrit case, and the improvement with DD is even more pronounced. Note that the difference between the DD sequences is much smaller than in the qutrit case, except for the $4\times 4X_4$ case at times $<25\ \mu$s. This could be attributed to the stronger $1/f$ dephasing suppression effect of the $4X_4$ sequences relative to the $3X_3$ sequences, which outweighs the accumulation of pulse errors. These ququart results highlight the effectiveness of DD in higher dimensions and its critical role in suppressing noise in more complex systems.

\begin{figure*}[t]
    \centering
    \includegraphics[width=0.45\linewidth]{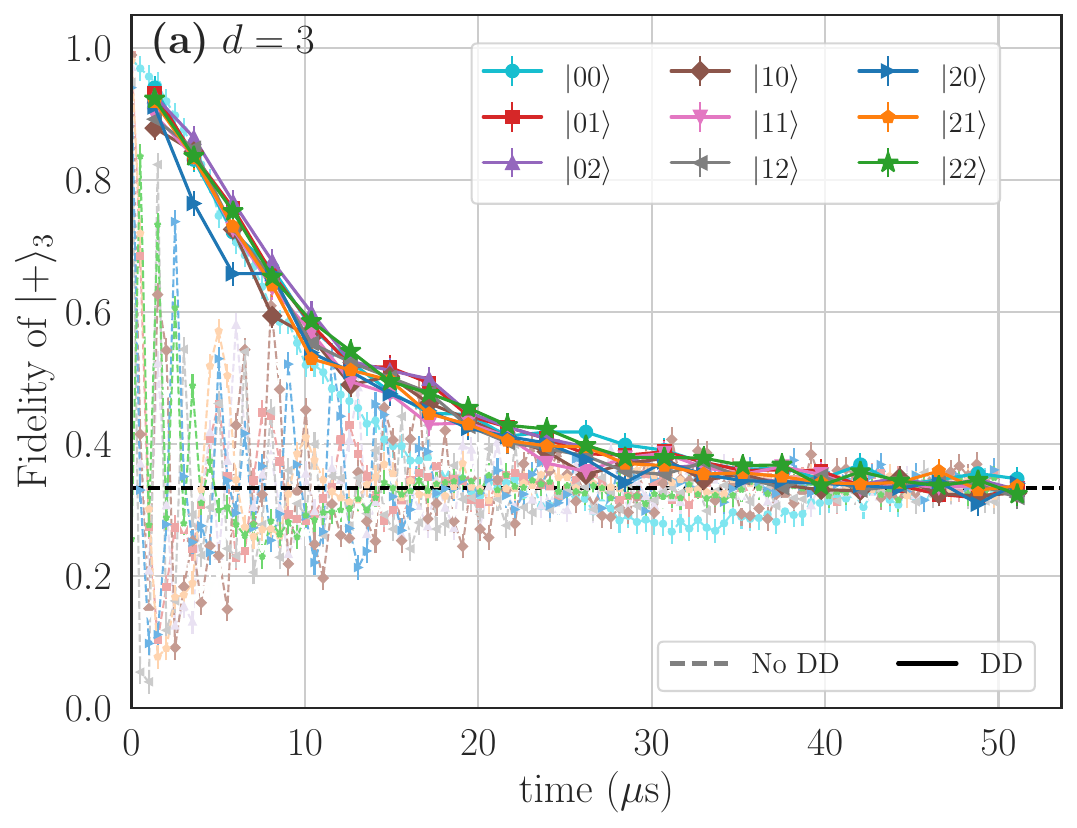}
    \includegraphics[width=0.45\linewidth]{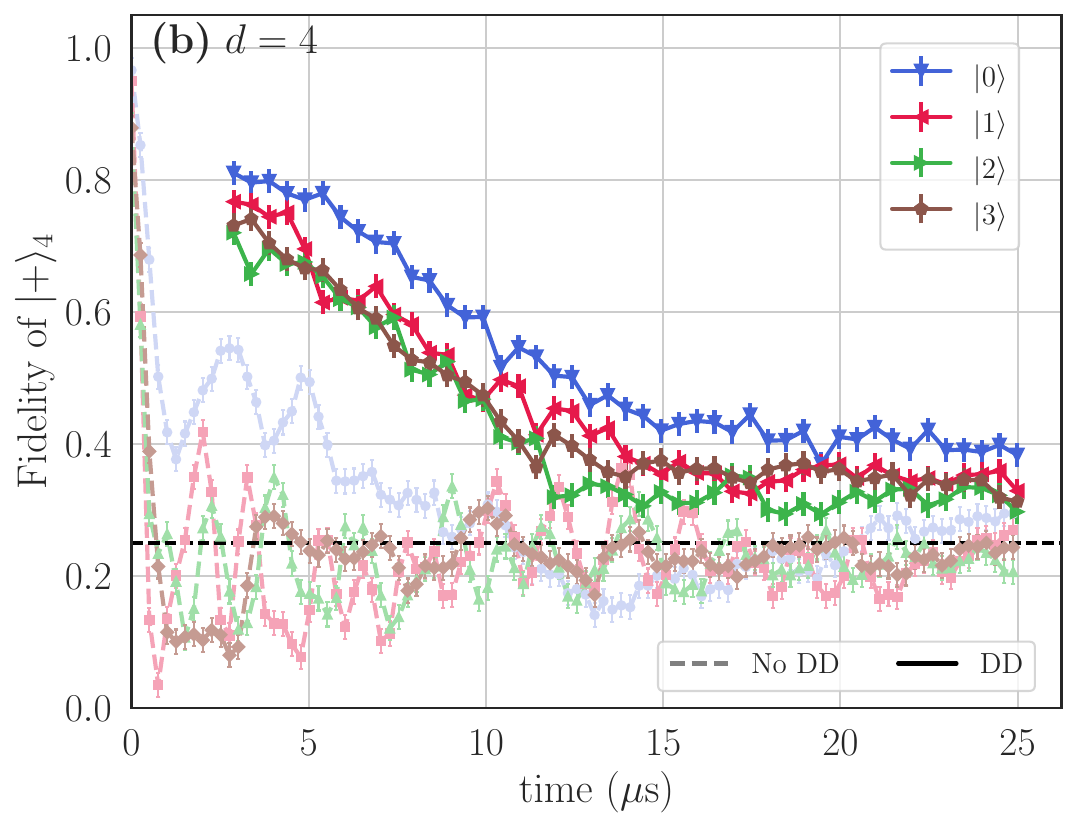}
    \caption{Experimental results showing suppression of cross-Kerr interactions using the CKDD sequence. (a) Fidelity of the qutrit superposition state $|+\rangle_3$, with CKDD (solid) and without (dashed). CKDD is a $9\tau$-long sequence, with $\tau_{\min} = 120$ ns. (b) Fidelity of the ququart superposition state $|+\rangle_4$, with CKDD (solid) and without (dashed). CKDD is a $16\tau$-long sequence, with $\tau_{\min} = 180$ ns. As in \cref{fig:single_qutrit}, longer evolution times correspond to a single repetition of CKDD with increased $\tau$. CKDD removes the cross-Kerr oscillations and improves the fidelity in both cases, converging to the fully mixed state fidelity baseline (dashed horizontal line) more slowly than the free evolution (No DD) curves. See text for further details.}
    \label{fig:spectator_qutrit}
\end{figure*}

\textit{Cross-Kerr suppressing DD} (CKDD).--- Having shown significant improvements with single-qudit DD against decoherence, we now deploy DD to mitigate qudit crosstalk. Recent work has demonstrated the efficacy of DD in suppressing coherent crosstalk errors in qubit systems~\cite{Tripathi2022, Zhou2023, evert2024syncopated}. In our fixed linear coupling qudit processor, single-qudit operations suffer from always-on crosstalk, which is a generalization of the $ZZ$ interaction between transmon qubits~\cite{Tripathi2019}. This type of crosstalk is commonly referred to as cross-Kerr interactions, which describe the spectator-state-dependent shifts of the relevant qudit transition frequencies. For two coupled qubits, in the lab frame where $H_{\rm 2q} = \sum_{i,j=0}^1 E_{ij}\ketb{ij}{ij}$ (in the eigenbasis), $\Tr(ZZ H_{\rm 2q})=E_{11}+E_{00} - E_{10} - E_{01}$. To model cross-Kerr interactions, consider the rotating frame that nullifies all bare transmon energy terms, leaving only the diagonal interaction terms. Then the Hamiltonian for the two-coupled transmon qudits simplifies to~\cite{Blok2021, Goss2022}:
\begin{align}
    H_{\rm CK} &=\sum_{i,j=1}^{d-1}\alpha_{ij} \ketb{ij}{ij} .
    \label{eq:crosskerr}
\end{align}
Here, $\alpha_{ij} = \omega_{ij}+\omega_{00} - \omega_{i0} - \omega_{0j}$ (taking $\hbar=1$), where $i,j \in \mathbb{Z}_d$, are the qudit frequency shifts (see \cref{fig:pulse}).

Although these cross-Kerr interactions, along with off-resonantly applied microwave drives, have been shown to facilitate entangling operations~\cite{Goss2022, Blok2021}, even minor cross-Kerr interactions during idle periods can introduce significant coherent errors. Building on the qudit DD formalism developed above, we now propose a DD sequence for coupled qudits. This DD sequence effectively suppresses all the cross-Kerr interactions, thereby enhancing system stability and operational fidelity.

The evolution operator due to $dX_d$ applied only to the first qudit is given by $U^{(1)}_{d\tau}\equiv dX_d\otimes I$. By concatenating this sequence with the same $dX_d$ sequence applied to the second qudit, we obtain the total evolution $U_{d^2\tau} \equiv U^{(2)}_{d\tau}\circ U^{(1)}_{d\tau}$, i.e.,
\begin{align}
U_{d^2\tau} = (I\otimes X_d) U^{(1)}_{d\tau} (I\otimes X_d) \cdots U^{(1)}_{d\tau}(I\otimes X_d) U^{(1)}_{d\tau} .
\label{eq:CKDD}
\end{align}
Concatenation of DD sequences was originally introduced in order to obtain high-order suppression~\cite{Khodjasteh:2005xu}; here, it serves the purpose of staggering the sequences on the two qudits, thus generalizing the idea of robust qubit-crosstalk suppression via staggering~\cite{Zhou2023}. We show in the SM that $U_{d^2\tau} = e^{i\theta} I\otimes I + O(T^2)$, where $\theta$ is a global phase, and $T= d^2\tau$. 
Thus, we expect that, to first order in the pulse interval, the $d^2\tau$-long \emph{cross-Kerr DD} (CKDD) sequence in \cref{eq:CKDD} suppresses all crosstalk between two coupled qudits. 

\begin{figure}[t!]
\includegraphics[width=.95\columnwidth]{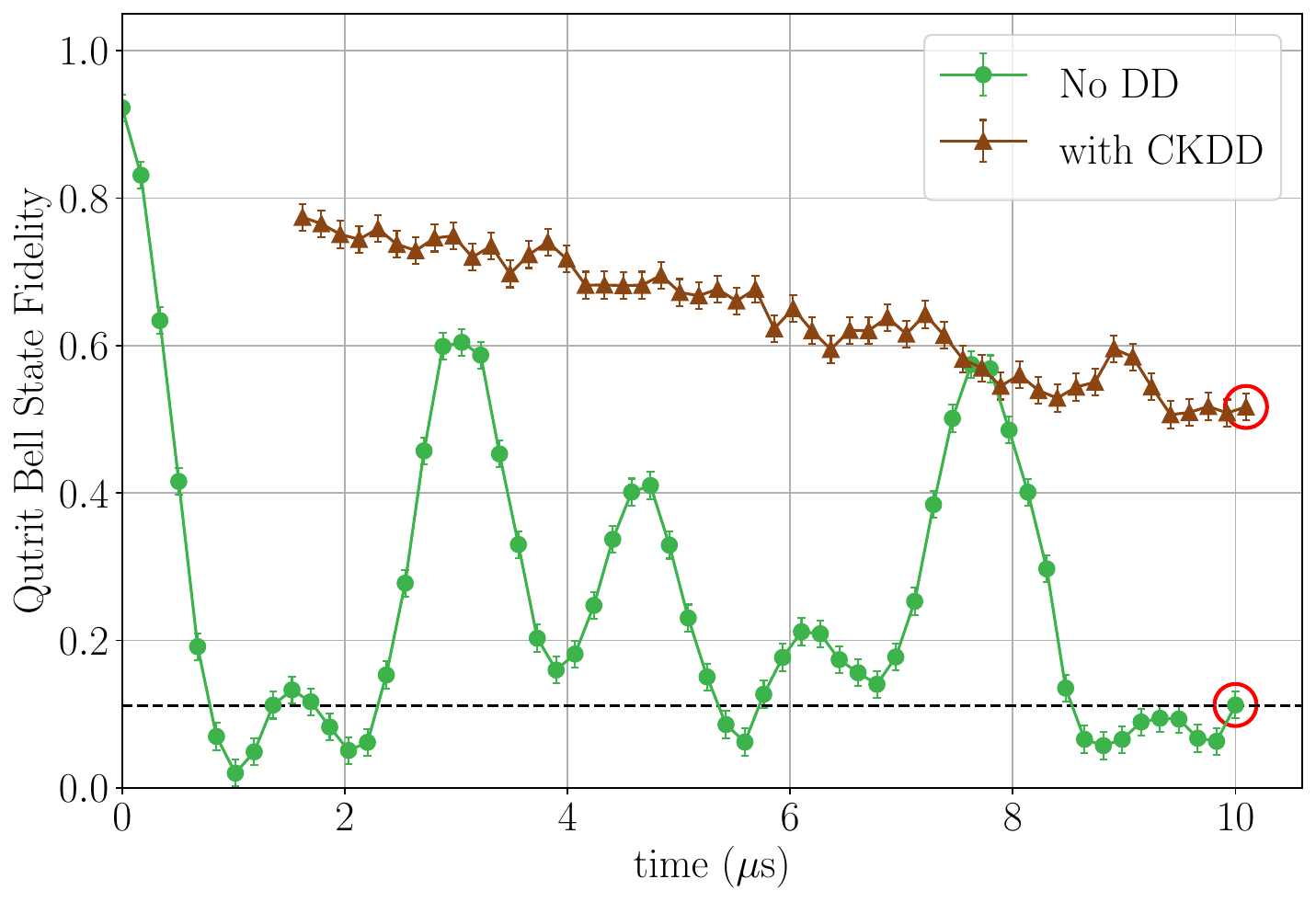}
\includegraphics[width=.95\columnwidth]{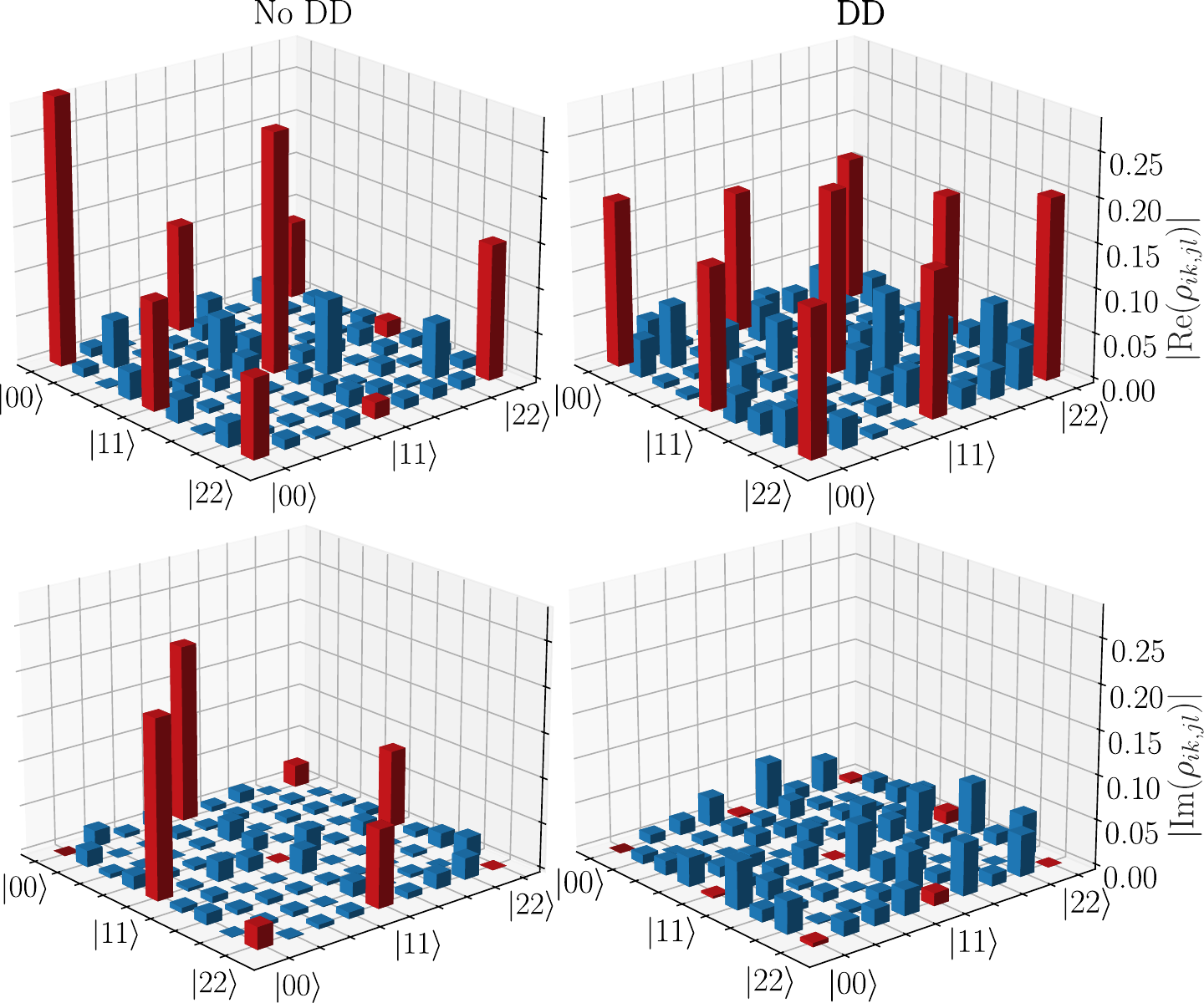}
\caption{Top: Fidelity of the qutrit Bell state over time with and without CKDD. See text for details. Bottom: real (upper) and imaginary (lower) components of quantum state tomography results for the final time point in the top plot (red circles). Ideally, $\Re(\rho_{ik,jl}) = \frac{1}{3}\delta_{ik}\delta_{jl}$ and $\Im(\rho_{ik,jl}) = 0$ for $i,j,k,l\in\mathbb{Z}_3$, where $\rho$ is the density matrix. Left: free evolution. The nine red-colored bars at the ideal positions are of varying magnitude and some contain large imaginary components, indicating deviations from the ideal qutrit Bell state. Right: with CKDD. In contrast, the nine red-colored bars are nearly uniform in height are have negligible imaginary components, indicating proximity to the ideal qutrit Bell state. Note that here CKDD is a $9\tau$-long sequence with $\tau_{\min} = 180~$ns, the $X_3$ gate duration; this differs from the previous figures where the $X_3$ gate duration is $120~$ns due to a different calibration.}
\label{fig:Bell_state}
\end{figure}

\textit{Experimental validation of CKDD}.--- 
To validate the CKDD sequence, we conduct experiments on coupled transmon qutrits and ququarts.  \cref{fig:spectator_qutrit}(a) shows the results for a linear chain of three coupled qutrits, where we prepare the set of nine initial states $\ket{i}\otimes\ket{+}_3\otimes\ket{j}$, $i,j\in\mathbb{Z}_3$. We trace out the states of the left and right ``spectator'' qutrits, and display the fidelity of the middle (main) qutrit's state with respect to $|+\rangle_3$ as a function of delay time. As depicted by the dashed light-color curves, different initial spectator states exhibit distinct curves that oscillate at different frequencies. The high frequency oscillations and the differences between the curves are attributable to the cross-Kerr interactions between the main qutrit and the two spectators. Next, we apply the CKDD sequence as $U^{(2)}_{d\tau}\circ (U^{(1)}_{d\tau}\otimes U^{(3)}_{d\tau})$, that is, an inner sequence where $dX_d$ is applied simultaneously to the spectators, and an outer sequence where $dX_d$ is applied to the main qutrit. The spectators' pulses are synchronized, so the sequence still takes a total time of $d^2\tau$. The resulting solid, bold-color curves exhibit a higher fidelity and none of variation of the free evolution curves, highlighting the efficacy of the CKDD sequence in suppressing cross-Kerr interactions and stabilizing the system dynamics for different initial spectator states. 

In \cref{fig:spectator_qutrit}(b), we present the fidelity results for two coupled ququarts, for the set of four initial states $\ket{i}\otimes \ket{+}_4$, $i\in\mathbb{Z}_4$ for the spectator. We again trace out the spectator state. Similarly to the qutrit case, under free evolution we observe different curves corresponding to different states of the spectator, but when CKDD is applied, all four curves exhibit a similar exponential decay, indicating suppression of the cross-Kerr interactions. The CKDD sequence consists of the ququart shift operators $X_4$ comprising three $\sigma^x_{i,i+1}$ gates in the two-level subspaces spanned by $\{\ket{i},\ket{i+1}\}$, $i\in\mathbb{Z}_4$. In the presence of large cross-Kerr interactions, driving the two-level subspaces is prone to large detuning errors. This results in somewhat lower CKDD fidelities for the ququart experiments compared to qutrits. Despite this, the suppression of the cross-Kerr interactions is clearly evident from the results, thus providing a proof-of-principle demonstration of the scalability of the CKDD protocol to higher dimensions.

\textit{Qutrit entanglement preservation via CKDD}.--- 
To demonstrate the effectiveness of CKDD beyond the preservation of product states, we prepare the qutrit Bell state $(\ket{00} + \ket{11} + \ket{22}/\sqrt{3}$  and measure its fidelity over time with and without CKDD. The preparation of this state involves a qutrit controlled-phase (CZ) gate~\cite{Goss2022}. We employ quantum state tomography to compute the fidelity of the Bell state. As depicted in \cref{fig:Bell_state}, the experimental results for free evolution (No DD) and with CKDD applied to both qutrits contrast sharply. In the absence of DD, the state fidelity suffers significantly due to strong coherent errors arising from large cross-Kerr interactions. The fidelity drops to near zero in $\sim 1\ \mu$s, meaning that the state evolves to an orthogonal qutrit Bell state, then oscillates around the fully mixed state fidelity baseline of $1/9$ (dashed horizontal line), highlighting the importance of suppressing cross-Kerr interactions. In contrast, with CKDD we observe a marked improvement; the oscillations are nearly eliminated, and the fidelity remains $>50\%$ even after $10~\mu$s. The state tomography histograms in \cref{fig:Bell_state} further highlight CKDD's ability to maintain the integrity of qutrit Bell states.

\textit{Conclusions and outlook}.---
Building on the theory of qudit DD~\cite{Wocjan2006, rotteler2006equivalence, Rotteler_Wocjan_2013, Bookatz_2016}, we have demonstrated the suppression of decoherence in transmon-based qutrits and ququarts. Our experimental results exhibit a substantial improvement in the preservation of the fidelity of superposition states of such qudit systems. Beyond decoherence, a significant challenge in scaling superconducting qudit systems arises from the persistent cross-Kerr interactions between coupled qudits. To address this, we introduced cross-Kerr DD as a protocol aimed at suppressing these spurious interactions. Our experimental results demonstrate that the CKDD sequence successfully suppresses cross-Kerr interactions in both qutrits and ququarts, which suggests that CKDD can be employed in higher-dimensional systems as well. Furthermore, we have shown that CKDD significantly improves the fidelity of maximally entangled qutrit states.  

Our findings broaden the scope of dynamical decoupling used in the service of the suppression of decoherence and crosstalk beyond the traditional setting of qubits to qudit systems. While our focus here was on transmons, our findings can be directly applied to other quantum computing platforms with access to qudits, such as fluxonium systems operated at half flux~\cite{nguyen2019high,nguyen2022blueprint}. This addition to the quantum noise suppression toolkit will hopefully benefit the development of scalable qudit-based quantum processors.

\textit{Acknowledgement---}
V.T., A.V. and D.A.L. were supported by the ARO MURI grant W911NF-22-S-0007 and by the National Science Foundation Quantum Leap Big Idea under Grant No. OMA-1936388. N.G. was supported by funding from the National Science Foundation under Grant No.~2210391. L.B.N. acknowledges funding from the Office of Advanced Scientific Computing Research, Testbeds for Science program, Accelerated Research in Quantum Computing Program, Office of Science of the U.S. Department of Energy under Contract No. DE-AC02-05CH11231.

\bibliography{refs.bib}

\clearpage

\appendix
\onecolumngrid
\begin{center}
\large \bf Supplemental Material: Qudit Dynamical Decoupling on a Superconducting Quantum Processor

\end{center}

\twocolumngrid

\section{Experimental Device Characterization}
\label{app:experimental_params}
The superconducting qudit device employed in this work consists of 8 fixed-frequency transmon qudits coupled together by coplanar waveguide resonators in a ring topology. For all the experiments in this work, we employ a susbset of the device consisting of a $3$-qudit line. We report the basic single qudit parameters of the subset of the device used in this work in Table.~\ref{tab:qudit_params}. For further, more extensive characterization of the device, including gate fidelities and readout fidelities, see Refs.~\cite{nguyen2023empowering,goss2023extending,Goss2022}.

\begin{table}[h!]
  \begin{center}
  \begin{tabular}
  {|c|c|c|c|}
  \hline
    Parameters & Q$_1$ & Q$_2$ & Q$_3$  \\[0.5ex]
    \hline
    \hline

    $\omega_{01}/2\pi$ (GHz) & 5.333 & 5.396 & 5.572   \\
    $\omega_{12}/2\pi$ (GHz) & 5.061 & 5.124 & 5.303 \\
    $\omega_{23}/2\pi$ (GHz) & 4.757 & 4.821 & 5.005 \\
    Avg. $T_1^{01}$ ($\mu$s)& 50(4) & 49(4) & 60(5) \\
    Avg. $T_1^{12}$ ($\mu$s)& 35(2) & 35(4) & 31(8) \\
    Avg. $T_1^{23}$ ($\mu$s)& 24(4) & 26(3) & 23(4) \\
    Avg. $T_{2e}^{01}$ ($\mu$s)& 78(5) & 85(9) & 90(6) \\
    Avg. $T_{2e}^{12}$ ($\mu$s)& 57(4) & 57(4) & 56(9) \\
    Avg. $T_{2e}^{23}$ ($\mu$s)& 26(2) & 27(2) & 24(3) \\
    \hline
    
  \end{tabular}
\end{center}
\caption{Transition frequencies $\omega_{ij}= (E_j-E_i)/\hbar$ up to $d=4$ of the qudits employed in our DD experiments. Additionally, we provide the two-level subspace mean $T_1$ and $T_2$ echo times of the device calculated from 100 repetitions of each coherence experiment.}
\label{tab:qudit_params}
\end{table}
$Q_2$ was used for both single-qudit DD experiments reported in \cref{fig:single_qutrit} of the main text.

\subsubsection{Qudit Cross-Kerr Coupling Rates}

As discussed in the main text, the fixed-linear coupling between superconducting qudits generates a longitudinal hybridization mediated largely by the higher levels present in each transmon well. These longitudinal interactions impart entangling phases to all qudit states $\ket{i,j}$ ($i,j \in \mathbb{Z}_d$) and lead to significant coherent errors.  In the doubly rotating frame of the two-qudit system, the effective cross-Kerr Hamiltonian defines the accumulation of all non-local entangling phases and is given by
%\begin{align}
    $H_{\rm CK} =\sum_{i,j=1}^{d-1}\alpha_{ij} \ketb{ij}{ij}$~\cite{Blok2021, Goss2022} [\cref{eq:crosskerr} in the main text].
%\label{eq:crosskerrSM}.
%\end{align}
In the cross-Kerr DD experiments reported in \cref{fig:spectator_qutrit} of the main text, we demonstrate that local pulses are sufficient to provide effective suppression of the $d=3$ and $d=4$ cross-Kerr interaction as well as of the system-bath coupling.

\subsubsection{$d=3$ cross-Kerr DD Experiment}
For our $d=3$ cross-Kerr DD experiment [see \cref{fig:spectator_qutrit}(a) of the main text], we prepared the $\ket{+}_3$ on $Q_2$ using a qutrit Hadamard gate and simultaneously prepared the two spectator qutrits $Q_1$ and $Q_3$ in states $\ket{i,j}$ ($i,j \in \mathbb{Z}_3$). We then allowed the system to time-evolve with and without our cross-Kerr DD sequence, and assessed the time-evolved state fidelity by remapping $\ket{+}_3$ to $\ket{0}$ via a final qutrit Hadamard gate. The relevant four qutrit cross-Kerr rates for this experiment measured via conditional Ramsey experiments are presented in \cref{tab:qutrit-ck}.

\begin{table}[h!]
  \begin{center}
  \begin{tabular}
  {|c|c|c|}
  \hline
    Parameters$/2\pi$ & $Q_1$-$Q_2$ & $Q_2$-$Q_3$  \\[0.5ex]
    \hline
    \hline

    $\alpha_{11}$ (MHz) & 0.112 & 0.212   \\
    $\alpha_{12}$ (MHz) & 0.623 & 0.465    \\
    $\alpha_{21}$ (MHz) & -0.515 & -0.162   \\
    $\alpha_{22}$ (MHz) & 0.341 & 0.615    \\

    \hline
    
  \end{tabular}
\end{center}
\caption{The qutrit cross-Kerr crosstalk rates present in the spectator DD experiment in \cref{fig:spectator_qutrit}(a) in the main text.}
\label{tab:qutrit-ck}
\end{table}

\subsubsection{$d=4$ cross-Kerr DD Experiment}

For our $d=4$ cross-Kerr DD experiment [see \cref{fig:spectator_qutrit}(b) of the main text], we prepared the $\ket{+}_4$ state on $Q_2$ using a ququart Hadamard gate and simultaneously prepared the spectator qudit $Q_1$ in $\ket{i}$ ($i \in \mathbb{Z}_4$). We then allowed the system to time-evolve with and without our cross-Kerr DD sequence, and assessed the time-evolved state fidelity by remapping $\ket{+}_4$ to $\ket{0}$ via a final ququart Hadamard gate. The relevant nine ququart cross-Kerr rates for this experiment measured via conditional Ramsey experiments are presented in \cref{tab:ququart-ck}.

\begin{table}[h!]
  \begin{center}
  \begin{tabular}
  {|c|c|}
  \hline
    Parameters$/2\pi$ & $Q_1$-$Q_2$  \\[0.5ex]
    \hline
    \hline

    $\alpha_{11}$ (MHz) & 0.112   \\
    $\alpha_{12}$ (MHz) & 0.623\\
    $\alpha_{13}$ (MHz) & 0.021\\
    $\alpha_{21}$ (MHz) & -0.515   \\
    $\alpha_{22}$ (MHz) & 0.341   \\
    $\alpha_{23}$ (MHz) & 0.730   \\
    $\alpha_{31}$ (MHz) & 0.226   \\
    $\alpha_{32}$ (MHz) & -0.442   \\
    $\alpha_{33}$ (MHz) & 0.345   \\

    \hline
    
  \end{tabular}
\end{center}
\caption{The ququart cross-Kerr crosstalk rates present in the spectator DD experiment in \cref{fig:spectator_qutrit}(b) in the main text.}\label{tab:ququart-ck}
\end{table}

\section{General theory of qudit dynamical decoupling} 
\label{app:HWbackground}

Consider the total time-independent Hamiltonian of a system coupled to a bath $H=H_S\otimes I_B+I_S\otimes H_B+H_{SB}$ where $H_S$, $H_B$, and $H_{SB}$ are the Hamiltonian terms associated with the system, the bath, and the system-bath interaction, respectively. Here, $H_S$ represents undesired system terms such as crosstalk and stray local fields. Pulses are applied to the system via an additional, time-dependent control Hamiltonian. Correspondingly, the \emph{decoupling group} $\mathcal{G}_d$ is defined by a set of unitary transformations $g_j$ acting purely on the system: $\mathcal{G}_d=\left\{g_0, \cdots, g_K\right\}$, where $g_0$ is the $d$-dimensional identity operator $I$.  Under the instantaneous (zero width) and ideal (error-free) pulse assumptions, cycling over all elements of the group yields the following DD pulse sequence~\cite{Zanardi:1999fk,Viola1999PRL}: 
\begin{equation} 
U(T)=
\prod_{j=0}^{|\mathcal{G}_d|-1} g_j^{\dagger} f_\tau g_j = e^{-iT H'} + \mathcal{O}(T^2).
\label{eq:commutant}
\end{equation}
Here, $\tau$ is the pulse interval (the time between consecutive pulses), $T=|\mathcal{G}_d|\tau$ is the total time taken by the sequence, and $f_\tau = e^{-i\tau H}$ is the free-evolution unitary. The effective Hamiltonian at the end of the sequence is
\beq
H' = H'_{SB} + H'_S + H_B ,
\eeq
where
\beq
H'_{SB} = \mathcal{P}_{\mathcal{G}_d}(H_{SB})\ , \quad H'_{S} = \mathcal{P}_{\mathcal{G}_d}(H_{S}) .
\eeq
Here
\beq
\mathcal{P}_{\mathcal{G}_d}(\Omega)= \frac{1}{|\mathcal{G}_d|}\sum_{j=0}^{|\mathcal{G}_d|-1} g_j^\dagger \Omega g_j
\eeq
is the projection of the operator $\Omega$ into the commutant of $\mathcal{G}_d$, i.e., the set of operators that commute with every element of $\mathcal{G}_d$. Crucially, this projection can be made proportional to $I$ or even vanish via a proper choice of $\mathcal{G}_d$. When $H'_{SB}=H'_S=0$, we call $\mathcal{G}_d$ and the corresponding DD sequence universal. For example, a universal DD sequence for a qubit ($d=2$) is obtained by choosing the decoupling group as the Pauli group $\mathcal{G}_2=\mathcal{P}$, which leads to the well-known XY4 sequence $U = Y f_\tau X f_\tau Y f_\tau X f_\tau$~\cite{Maudsley:1986ty}. 

For $d\ge 2$, we instead use the corresponding Heisenberg-Weyl group (HWG) of order $d^2$, which generalizes the Pauli group. We define shift and phase operators as in \cref{eq:fund-ops} of the main text, repeated here for convenience:
\begin{equation} 
    X_d\equiv \sum_{k=0}^{d-1} \ketb{(k+1)\bmod d}{k}, \, 
    \quad Z_d\equiv \sum_{k=0}^{d-1} \gamma_d^k \ketb{k}{k},
\label{eq:fund-ops-SM}
\end{equation}
where $\gamma_d=e^{2\pi i/d}$ is the $d$'th root of unity.

Note that $X_d^d= Z_d^d = I$, and that $X_d$ and $Z_d$ are generally non-Hermitian but are both unitary for all $d$ and hence satisfy
\beq
(X_d^\dag)^\alpha X_d^\beta = X_d^{\beta-\alpha} , \quad (Z_d^\dag)^\alpha Z_d^\beta = Z_d^{\beta-\alpha} .
\label{eq:XdagZdag}
\eeq
In particular, 
\beq
X_d^\dag = X_d^{-1} = X_d^{d-1}\ , \quad Z_d^\dag = Z_d^{-1} = Z_d^{d-1}.
\eeq
$X_d$ and $Z_d$ are the generators of the HWG, whose elements are 
\begin{equation}
    \Lambda_{\alpha\beta}=(-\sqrt{\gamma_d})^{\alpha\beta}X_d^\alpha Z_d^\beta,
    \label{eq:Lambda}
\end{equation}
where $\alpha,\beta\in\mathbb{Z}_d$. For $d=2$, the HWG trivially reduces to the Pauli group.

For the two generators $X_d$ and $Z_d$, we have 
\bes
\begin{align}
X_d Z_d &= \sum_{k=0}^{d-1}e^{2\pi i k/d} \ketb{k+1\bmod d}{k}\\
Z_d X_d &= \sum_{k=0}^{d-1} e^{2\pi i (k+1)/d}\ketb{k+1\bmod d}{k} ,
\end{align}
\ees
i.e., 
%using $\gamma_d^{d-1} = e^{-2\pi i/d} = \gamma_d^{-1}$:
\beq
\label{eq:hw-comm-prop-basic}
Z_d X_d=\gamma_d X_d Z_d.
\eeq
Similarly, we find:
\beq
\label{eq:hw-comm-prop-basic2}
Z_d^\dag X_d=\gamma_d^{-1} X_d Z_d^\dag.
\eeq
Using \cref{eq:hw-comm-prop-basic,eq:hw-comm-prop-basic2}, we can show that
\bes
\label{eq:hw-comm-prop}
\begin{align} 
\label{eq:hw-comm-prop-1}
     Z_d^\beta X_d^\alpha &=\gamma_d^{\alpha\beta}  X_d^\alpha Z_d^\beta\\
\label{eq:hw-comm-prop-2}
     (Z_d^\dag)^\beta X_d^\alpha &=\gamma_d^{-\alpha\beta} X_d^\alpha (Z_d^\dag)^\beta .
\end{align}
\ees

\begin{proof}
For $\alpha =\beta = 1$, \cref{eq:hw-comm-prop} reduces to \cref{eq:hw-comm-prop-basic}.
Consider $\beta \ge 2$:
\beq
   Z_d^\beta X_d = \gamma_d Z_d^{\beta-1} X_d Z_d = \dots =\gamma_d^\beta X_d Z_d^\beta .
\eeq
When $\alpha \ge 2$:
\bes
\begin{align}
   Z_d^\beta X_d^\alpha &= (Z_d^\beta X_d) X_d^{\alpha-1} = \gamma_d^\beta X_d (Z_d^\beta X_d) X_d^{\alpha-2} \\
   &= \gamma_d^{2\beta} X_d^2 Z_d X_d^{\alpha-2} =\dots = \gamma_d^{\alpha\beta}X_d^\alpha Z_d^\beta .
\end{align}
\ees
\cref{eq:hw-comm-prop-2} follows analogously.
\end{proof}
This means that the commutation relations for two arbitrary HW operators are
\bes
\label{eq:comm-dagg}
\begin{align}
\label{eq:comm-dagg-1}
\Lambda_{\alpha\beta}\Lambda_{\mu\nu} &=
\gamma_d^{\beta\mu-\alpha\nu}\Lambda_{\mu\nu}\Lambda_{\alpha\beta} \\
\label{eq:comm-dagg-2}
\Lambda_{\alpha\beta}^\dag\Lambda_{\mu\nu} &= \gamma_d^{\alpha\nu-\beta\mu}\Lambda_{\mu\nu}\Lambda_{\alpha\beta}^\dag . 
\end{align}
\ees
Unless $\alpha=\beta=0$, the HW operators are non-Hermitian for $d\ge 3$
\bes
\label{eq:non-hermitian} 
\begin{align}
    \Lambda_{\alpha\beta}^{\dag} &= e^{-i\pi \alpha\beta\frac{d+1}{d}}(X_d^\alpha Z_d^\beta)^\dag \\
    &= (-\sqrt{\gamma_d})^{-\alpha\beta}(Z_d^\dag)^\beta (X_d^\dag)^\alpha \ne \Lambda_{\alpha\beta} ,
\end{align}
\ees
but unitary for all $d$:
\beq
\Lambda_{\alpha\beta}^{\dag} \Lambda_{\alpha\beta} = (Z_d^\dag)^\beta (X_d^\dag)^\alpha X_d^\alpha Z_d^\beta = I ,
\eeq
where we used the unitarity of $X_d$ and $Z_d$. Combining unitarity with \cref{eq:comm-dagg-1}, we obtain the identity
\beq
\Lambda_{\alpha\beta}^{\dag} \Lambda_{\mu\nu}\Lambda_{\alpha\beta} = \gamma_d^{\alpha\nu-\beta\mu}\Lambda_{\mu\nu}  ,
\label{eq:HW-group-3ops}
\eeq
which will prove to be crucial below for demonstrating that the HWG is a universal DD group.

The operators $\{\Lambda_{\alpha\beta}\}_{\alpha,\beta\in\mathbb{Z}_d}$ form an irreducible, unitary, and projective representation of the HWG over the $d$-dimensional system Hilbert space when $d$ is a prime power ($d=p^k$ for prime number $p$ and positive integer $k$). This implies, by Schur's Lemma, that both $H'_{SB}$ and $H'_S$ are proportional to $I$ or vanish. Therefore, the unitary $U(T)$ defined in \cref{eq:commutant} reduces (up to a global phase) to the identity operation on the system, i.e., the condition for first-order decoupling is satisfied. The latter (or an equivalent one using group character tables) was the argument used in Refs.~\cite{Wocjan2006, rotteler2006equivalence, Rotteler_Wocjan_2013, Bookatz_2016}; going beyond the case of prime powers, we now show that, in fact, first-order decoupling holds for arbitrary $d$.

The HW operators also form an operator basis for the $d$-dimensional system Hilbert space. Thus, we can expand $H = H_S\otimes I_B + H_{SB} + I_S\otimes H_B$ as 
\beq
H = \sum_{\mu,\nu=0}^{d^2-1}\Lambda_{\mu\nu}\otimes B_{\mu\nu} ,
\label{eq:H-HW}
\eeq 
where $B_{\mu\nu}$ is either zero, proportional to $I_B$ (to account for $H_S\otimes I_B$), or is a non-identity bath operator. The term with $\mu=\nu=0$ corresponds to the pure-bath term $I_S\otimes H_B$.

Now recall that cycling over the decoupling group $\mathcal{G}_d$ yields \cref{eq:commutant}. Choosing the decoupling group as the HWG $\{\Lambda_{\alpha\beta}\}$ means that the effective Hamiltonian becomes 
\bes
\label{eq:H'}
\begin{align}
\label{eq:H'1}
   H' &= \mathcal{P}_{\mathcal{G}_d}(H) \\
   &= \frac{1}{d^2} \sum_{\alpha,\beta=0}^{d^2-1} \Lambda_{\alpha\beta}^{\dag} \sum_{\mu,\nu=0}^{d^2-1}\Lambda_{\mu\nu}\Lambda_{\alpha\beta}\otimes B_{\mu\nu} \\
\label{eq:H'2}
    &= \frac{1}{d^2} \sum_{\mu,\nu=0}^{d^2-1}f_{\mu\nu}\Lambda_{\mu\nu}\otimes B_{\mu\nu} ,
\end{align}
\ees
where, using \cref{eq:HW-group-3ops},
\beq
f_{\mu\nu} =\sum_{\alpha,\beta=0}^{d^2-1} \gamma_d^{\alpha\nu-\beta\mu}  .
\eeq
Let us now show that 
\beq
f_{\mu\nu} = d^4\delta_{\mu0}\delta_{\nu0} .
\label{eq:fmunu}
\eeq
Intuitively, this follows from the zero-sum property of the roots of unity: $\sum_{k=0}^{d-1}\gamma_d^k=0$.

\begin{proof}
Note that $f_{\mu\nu}= h^*_{\mu}h_\nu$, and 
\bes
\begin{align}
    h_\nu &= \sum_{\alpha=0}^{d^2-1} \gamma_d^{\nu\alpha}  =  \sum_{k=0}^{d-1} \sum_{j=0}^{d-1} e^{2\pi i(kd+j)\nu/d} \\
    &= \sum_{k=0}^{d-1} \sum_{j=0}^{d-1} (e^{2\pi i \nu/d})^j = d S,
\end{align}
\ees
where $S = \sum_{j=0}^{d-1} \omega^j$ and $\omega=e^{2\pi i \nu/d}$.
If $\nu\ne 0$ then $\omega\ne 1$ is a $d$'th root of unity (since $\omega^d=1$). Multiplying both sides by $\omega - 1$ yields $(\omega - 1)S = \sum_{j=0}^{d-1} \omega^{j+1} - \sum_{j=0}^{d-1} \omega^j$.
The terms $\omega, \omega^2, \ldots, \omega^{d-1}$ cancel out, leaving
$(\omega - 1)S = \omega^d - 1 = 0$.
Since $\omega \neq 1$ we can divide both sides by $\omega - 1$, giving $S = 0$. If $\nu=0$ then $S=d$. 
\end{proof}

Combining \cref{eq:H',eq:fmunu}, we finally have
\beq
H' = I_S\otimes H_B ,
\eeq
i.e., $H'_{SB}=H'_S=0$, leaving only the pure-bath term.
This proves that the HWG is a universal decoupling group.

\begin{figure}[t]
\includegraphics[scale=.48]{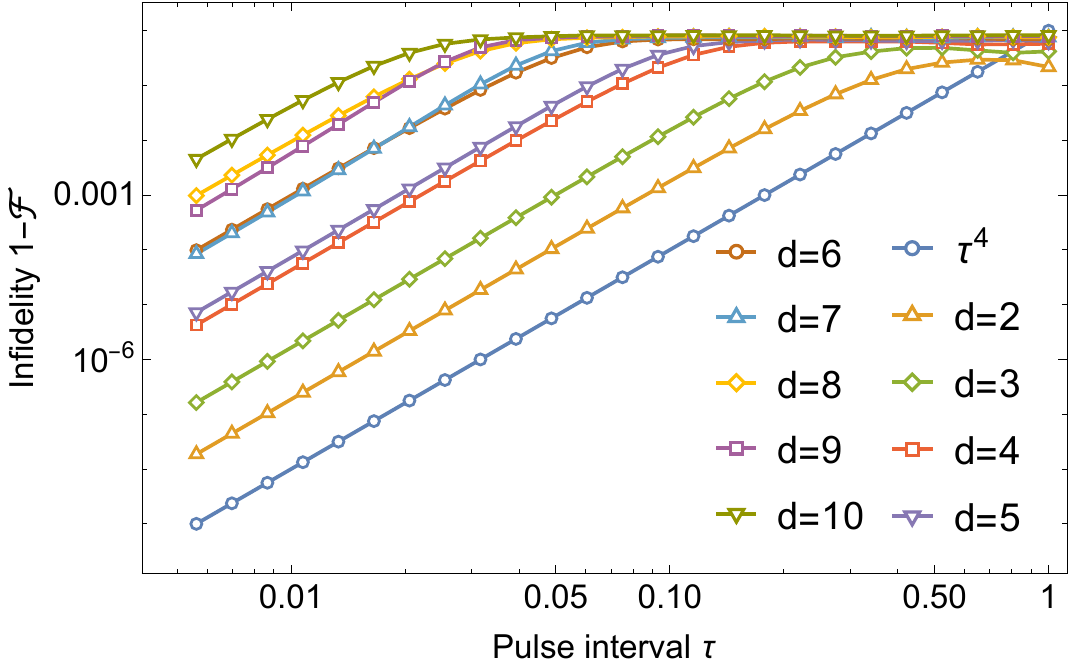}
\caption{Numerical simulation of the the universality of the DD sequence generated by cycling over the HWG for $2 \leq d\leq 10$. Here we plot the infidelity of the resulting unitary evolution as a function of the pulse interval time $\tau$. Since the sequence is expected to cancel the errors to the first order $\mathcal{O}(\tau^2)$, the infidelity should scale as $\mathcal{O}(\tau^4)$, as confirmed by our simulations.} 
\label{fig-arbitrary-d}
\end{figure}

We numerically confirm this universality in \cref{fig-arbitrary-d}. For various dimensions $2\leq d \leq 10$, we consider a system-bath Hamiltonian containing all the HW operators with randomized coefficients (i.e., a classical bath). We then apply the corresponding universal sequence and compare the fidelity of the resulting unitary to the identity operator $I$ in each case. Given that the errors in the unitary evolution under DD are suppressed to the first order [\cref{eq:commutant}], i.e., leaving the leading order term $\mathcal{O}(T^2)$ where $T\propto \tau$, we expect the fidelity to scale as $\mathcal{O}(\tau^4)$. This is confirmed in \cref{fig-arbitrary-d}. 

%%%%%%%%%%%%%%%%%%%%%%%%%%%%%%%%%
%%%%%%%%%%%%%%%%%%%%%%%%%%%%%%%%%

\section{Application to single-axis, pure dephasing noise} 
\label{app:single-axis}

We utilize the formalism developed above to analyze the single-axis noise problem (i.e., pure dephasing) as a special case. This is the basis for the transmon-based qutrit and ququart experiments we present in the main text, where dephasing is the dominant source of decoherence. In this single-axis scenario, the system-bath interaction component of \cref{eq:H-HW} reduces to
\begin{equation}
    H_{SB}^Z=\sum_{\nu=1}^{d-1}\Lambda_{0\nu}\otimes B_{\nu},
    \label{eq:HSB^Z}
\end{equation}
where $\Lambda_{0\nu} = Z_d^\nu$ [\cref{eq:Lambda}].

We could choose the full HWG as a decoupling group, but since the $Z_d$-type  HW operators commute with $H_{SB}^Z$, we need only consider the pure $X_d$-type decoupling operators $\Lambda_{\alpha 0} = X_d^\alpha$, which satisfy the commutation relations \cref{eq:comm-dagg} non-trivially. I.e., for pure qudit dephasing the relevant decoupling group is the order-$d$ HWG subgroup $\mathcal{G}_d^X = \{\Lambda_{\alpha 0}\}_{\alpha=0}^{d-1} = \{I, X_d, X_d^2, \cdots, X_s^{d-1}\}$.

To show that $\mathcal{G}_d^X$ decouples $H_{SB}^Z$, we observe, using \cref{eq:HW-group-3ops}, that the effective Hamiltonian is
\bes
\label{eq:H'Z}
\begin{align}
\label{eq:H'1Z}
   (H_{SB}^{Z})' &= \mathcal{P}_{\mathcal{G}^X_d}(H_{SB}^Z) \\
   &=\frac{1}{d} \sum_{\alpha=0}^{d-1} \Lambda_{\alpha 0}^{\dag} \sum_{\nu=1}^{d-1}\Lambda_{0\nu}\Lambda_{\alpha 0}\otimes B_{\nu} \\
\label{eq:H'2Z}
    &= \frac{1}{d}\sum_{\nu=1}^{d-1}f_\nu\Lambda_{0\nu}\otimes B_{\nu} ,
\end{align}
\ees
where
\beq
f_\nu = \sum_{\alpha=0}^{d-1}\gamma_d^{\nu \alpha} = d\delta_{\nu 0} ,
\label{eq:f_nu}
\eeq
and the last equality is again due to zero-sum property of the $d$'th root of unity. It follows that $(H_{SB}^{Z})'=0$, i.e., $\mathcal{G}_d^X$ is a decoupling group for qudit dephasing.

\section{Qutrit ($d=3$) dynamical decoupling and $3X_3$}

For concreteness, we now illustrate the results above by giving the explicit form of the HW subgroup $\mathcal{G}_d^X$ for the case of qutrit dephasing. This is the simplest non-trivial example going beyond qubits.

The generators of the qutrit HWG are $X_3$ and $Z_3$, which can be seen as generalizations of the Pauli matrices $\sigma_x$ and $\sigma_z$, respectively. Setting $\omega \equiv \gamma_3= e^{2\pi i/3}$ (the cube root of unity), the shift and phase operators are
\bes
\begin{align}
&    \Lambda_{10}=X_3 = \begin{pmatrix}
0 & 0 & 1 \\
1 & 0 & 0 \\
0 & 1 & 0
\end{pmatrix}\ ,\quad \Lambda_{01}=Z_3 = \begin{pmatrix}
1 & 0 & 0 \\
0 & \omega & 0 \\
0 & 0 & \omega^2
\end{pmatrix}\\
&    \Lambda_{20}=X_3^2 = \begin{pmatrix}
 0 & 1 & 0 \\
 0 & 0 & 1 \\
 1 & 0 & 0 \\
\end{pmatrix}\ ,\quad \Lambda_{02}=Z_3^2 = \begin{pmatrix}
1 & 0 & 0 \\
0 & \omega^2 & 0 \\
0 & 0 & \omega
\end{pmatrix}.
\end{align}
\ees
Their action on the qutrit computational basis states $|m\rangle$ ($m=0,1,2$) is $X_3|m\rangle = |(m + 1)\bmod 3\rangle$ and $Z_3|m\rangle = \omega^m|m\rangle$. Note that $O^{-1} = O^{\dagger} = O^2$ and $O^3=I$ for $O=X_3,Z_3, X_3^2$, and $Z_3^2$.

Their commutation properties follow from \cref{eq:hw-comm-prop}:
\bes
\begin{align}
X_3 Z_3 = \omega^2 Z_3X_3, &\quad  X_3^2 Z_3 = \omega Z_3X_3^2,\\
X_3 Z_3^2 = \omega Z_3^2 X_3, &\quad X_3^2 Z_3^2 = \omega^2
 Z_3^2X_3^2.
\end{align}
\ees
Since $X_3^\dagger=X_3^2$, we can write 
\bes
\begin{align}
(X_3^2)^\dagger Z_3 X_3^2 = \omega^2 Z_3, &\quad  X_3^\dagger Z_3 X_3 = \omega Z_3,\\
(X_3^2)^\dagger Z_3^2 X_3^2 = \omega Z_3^2 , &\quad X_3^\dagger Z_3^2 X_3 = \omega^2
 Z_3^2,
\end{align}
\ees
which is a special case of \cref{eq:HW-group-3ops}. The decoupling group is $\mathcal{G}_3^{X_3}=\{I, X_3, X_3^2\}$, which suppresses dephasing due to the system-bath interaction $H_{SB}^Z$ with $d=3$ [\cref{eq:HSB^Z}]. For example, consider free evolution subject to a term of the form $Z_3\otimes B$; writing out the DD sequence \cref{eq:commutant} explicitly, we have, for $T=3\tau$:
\bes
\label{eq:qutrit}
\begin{align} 
U(T) &=\big(I f_\tau I\big)\big((X_3^2)^\dagger f_\tau X_3^2\big)\big(X_3^\dagger f_\tau X_3\big) \\
&= e^{-i\tau Z_3\otimes B}e^{-i\tau (X_3^2)^\dagger Z_3 X_3^2 \otimes B}e^{-i\tau X_3^\dagger Z_3 X_3 \otimes B} \\
&= e^{-i\tau(Z_3 \otimes B)}e^{-i\tau  (\omega^2 Z_3 \otimes B)} e^{-i\tau (\omega Z_3 \otimes B)} \\
\label{eq:qutrit-d}
&= e^{-i\tau(1+\omega^2 + \omega)Z_3\otimes B} + \mathcal{O}(T^2) \\
&= I + \mathcal{O}(T^2),
\end{align}
\ees
i.e., suppression to second order of dephasing due to $Z_3\otimes B$. Replacing $Z_3\otimes B$ with $Z_3^2\otimes B'$ simply rearranges the order of the roots of unity, yielding $1+\omega + \omega^2 = 0$ instead of $1+\omega^2 + \omega = 0$ in \cref{eq:qutrit-d}, with the same outcome.

Note that since $(X_3^2)^\dagger = X_3 = X_3^2 X_3^\dagger$, this DD sequence in fact reduces to three equidistant pulses of type $X_3$: 
\beq
U(T) = X_3 f_{\tau} X_3 f_{\tau} X_3 f_{\tau} 
\eeq
which is the reason we called it $3X_3$ in the main text. We could have equivalently used the sequence consisting of three $X_3^2$ pulses.

\section{Proof of first order suppression of cross-Kerr coupling by the CKDD sequence}

The goal of the CKDD sequence is to suppress the cross-Kerr coupling Hamiltonian \cref{eq:crosskerr}:

\beq
    H_{\rm CK} =\sum_{i,j=1}^{d-1}\alpha_{ij} \ketb{i}{i}\otimes \ketb{j}{j}.
    \label{eq:crosskerr_supp}
\eeq
Since the HWG is an operator basis, and in particular its phase elements $\{Z_d^\alpha\}_{\alpha=0}^{d-1}$ are a basis for diagonal operators, we can expand each diagonal term in $H_{\rm CK}$ as
\beq
\ketb{i}{i} = \sum_{k=0}^{d-1} c_{ik} Z_d^k ,
\eeq
where
\beq
c_{ik} = \frac{1}{d}\Tr[(Z_d^\dag)^k\ketb{i}{i}] = \frac{1}{d}\gamma_d^{-ik}.
\eeq
This is a consequence of \cref{eq:XdagZdag} and the zero-sum property: $\Tr[(Z_d^\dag)^k Z_d^{l}] = \Tr(Z_d^{l-k}) = \sum_{m=0}^{d-1} \gamma_d^{m(l-k)} = d\delta_{kl}$. Thus, we can rewrite the cross-Kerr Hamiltonian as
\beq
H_{\rm CK} = \sum_{k,l=0}^{d-1} \zeta_{kl}Z_d^k\otimes Z_d^l,
\eeq
where 
\beq
\zeta_{kl} = \sum_{i,j=1}^{d-1} c_{ik} c_{jl} \alpha_{ij} = \frac{1}{d^2}\sum_{i,j=1}^{d-1}\gamma_d^{-(ik+jl)}\alpha_{ij} .
\eeq

The CKDD sequence $U_{d^2\tau}$ [\cref{eq:CKDD}] consists of an inner $X_d$-type sequence applied to the first qudit and an outer $X_d$-type sequence applied to the second qudit.
Namely, $U_{d^2\tau} \equiv U^{(2)}_{d\tau}\circ U^{(1)}_{d\tau} = (I_d\otimes X_d) U^{(1)}_{d\tau} (I_d\otimes X_d) \cdots U^{(1)}_{d\tau}(I_d\otimes X_d) U^{(1)}_{d\tau}$, where $U^{(1)}_{d\tau}\equiv dX_d\otimes I_d$ 
and $U^{(2)}_{d\tau}\equiv I_d\otimes dX_d$. 

Generalizing from the single qudit dephasing case, the effect of the inner $U^{(1)}_{d\tau}$ sequence is to project $H_{\rm CK}$ into the commutant of $\mathcal{G}^X_d\otimes I_d$, i.e.,
\bes
\label{eq:H'CKinner}
\begin{align}
   H'_{\rm CK} &= \mathcal{P}_{\mathcal{G}^X_d\otimes I_d}(H_{\rm CK}) \\
   &= \frac{1}{d}\sum_{\alpha=0}^{d-1} X_d^{\alpha\dag}\otimes I_d \sum_{k,l=0}^{d-1}\zeta_{kl}Z_d^k X_d^{\alpha}\otimes Z_d^l\\
    &= \frac{1}{d}\sum_{k,l=0}^{d-1}f_k \zeta_{kl} Z_d^k\otimes Z_d^l ,
\end{align}
\ees
where $f_k = \sum_{\alpha=0}^{d-1}\gamma_d^{k \alpha} = d\delta_{k 0}$, just as in \cref{eq:f_nu}. Thus, the effect of the inner sequence is to leave just the identity term on the first qudit:
\beq
H'_{\rm CK} = \sum_{l=0}^{d-1} \zeta_{0l} I_d \otimes Z_d^l .
\eeq
The outer sequence then removes the remaining dephasing terms:
\bes
\label{eq:H'CKouter}
\begin{align}
   H''_{\rm CK} &= \mathcal{P}_{I_d\otimes\mathcal{G}^X_d}(H'_{\rm CK}) \\
   &=\frac{1}{d}\sum_{\alpha=0}^{d-1} I_d\otimes X_d^{\alpha\dag} \sum_{l=0}^{d-1}\zeta_{0l} I_d \otimes Z_d^l X_d^{\alpha}\\
    &= \frac{1}{d}\sum_{l=0}^{d-1}f_l \zeta_{0l} I_d \otimes Z_d^l 
    = \zeta_{00} I_{d^2} .
\end{align}
\ees
Consequently, it follows from \cref{eq:commutant} that $U_{d^2\tau} = e^{-i \zeta_{00} T}I_d\otimes I_d + \mathcal{O}(T^2)$, where $T=d^2\tau$ and $\zeta_{00}=\frac{1}{d^2}\sum_{i,j=1}^{d-1}\alpha_{ij}$. This proves that the CKDD sequence achieves first-order decoupling of the cross-Kerr interaction.

Note that applying simultaneous $X_d$-type sequences to both qudits does not work. I.e., using $\mathcal{G}^X_d\otimes \mathcal{G}^X_d$ as the decoupling group results instead of \cref{eq:H'CKinner} in the projection 
\bes
\label{eq:H'CKsym}
\begin{align}
   H'_{\rm CK} &= \mathcal{P}_{\mathcal{G}^X_d\otimes \mathcal{G}^X_d}(H_{\rm CK}) \\
   &= \frac{1}{d^2}\sum_{\alpha=0}^{d-1} \sum_{k,l=0}^{d-1}\zeta_{kl}X_d^{\alpha\dag}Z_d^k X_d^{\alpha}\otimes X_d^{\alpha\dag}Z_d^l X_d^{\alpha\dag}\\
    &= \frac{1}{d^2}\sum_{k,l=0}^{d-1} \zeta_{kl} Z_d^k\otimes Z_d^l \sum_{\alpha=0}^{d-1}\gamma_d^{(k+l)\alpha} ,
\end{align}
\ees
and $\sum_{\alpha=0}^{d-1}\gamma_d^{(k+l)\alpha} = d\delta_{k+l,d}$, i.e., terms of the form $Z_d^k\otimes Z_d^{d-k}$ are not suppressed. This is a generalization of the qubit case, where it is well known that simultaneous $X$-type sequences do not cancel crosstalk \cite{Tripathi2022}. 

\end{document}